\documentclass[12pt]{article}

\usepackage{graphicx}
\usepackage{amsfonts}
\usepackage{amsthm}
\usepackage{amsmath}
\usepackage[top=3cm, bottom=3.25cm, left=2.75cm, right=2.75cm]{geometry}

\newtheorem*{theorem}{Theorem}

\newcommand{\Section}[1]{\section{#1} \setcounter{equation}{0}}
\newcommand{\be}{\begin{eqnarray}}
\newcommand{\ee}{\end{eqnarray}}
\newcommand{\beq}{\begin{equation}}
\newcommand{\eeq}[1]{\label{#1}\end{equation}}
\newcommand{\ber}{\begin{eqnarray}}
\newcommand{\eer}[1]{\label{#1}\end{eqnarray}}

\newcommand{\ie}{\textit{i.e.},~}

\def\+{{+\!\!\!+}}

\newcommand{\kah}{K\"ahler~}

\newcommand{\eg}{\textit{e.g.},~}
\newcommand{\egg}{\textit{e.g.}~}

\begin{document}
\renewcommand{\theequation}{\thesection.\arabic{equation}}
\setcounter{page}{0}
\thispagestyle{empty}

\begin{flushright} \small
IMPERIAL-TP-2008-CH-03 \\ UUITP-19/08  \\  YITP-SB-08-36\\Nordita-2008-48\\
\end{flushright}
\smallskip
\begin{center} \LARGE
{\bf  Generalized \kah geometry and gerbes}
\\[12mm] \normalsize
{\bf Chris M.~Hull$^{a}$, Ulf~Lindstr\"om$^{b}$, Martin Ro\v cek$^{c}$, \\
Rikard von Unge$^{c,d,e}$, and Maxim Zabzine$^{b}$} \\[8mm]
{\small\it
$^a$The Blackett Laboratory, Imperial College London\\
Prince Consort Road, London SW7 2AZ, U.K.\\
~\\
$^b$Department of Theoretical Physics
Uppsala University, \\ Box 803, SE-751 08 Uppsala, Sweden \\
~\\
$^c$C.N.Yang Institute for Theoretical Physics, Stony Brook University, \\
Stony Brook, NY 11794-3840,USA\\
~\\
$^d$Simons Center for Geometry and Physics, Stony Brook University, \\
Stony Brook, NY 11794-3840,USA\\
~\\
$^{e}$Institute for Theoretical Physics, Masaryk University, \\
61137 Brno, Czech Republic \\~\\}
\end{center}
\vspace{10mm}
\centerline{\bf\large Abstract}
\bigskip
\noindent
We introduce and study the notion of a biholomorphic gerbe with connection. 
The biholomorphic gerbe provides a natural geometrical framework for generalized \kah
geometry in a manner analogous to the way a holomorphic line bundle is related to \kah
geometry. The relation between the gerbe and the generalized \kah potential is discussed. 
\eject
\footnotesize
\tableofcontents{}
\normalsize
\eject

 \Section{Introduction}
Recently, generalized \kah geometry \cite{gualtieriPhD} 
has attracted considerable interest in the physics 
and mathematics communities. It was discovered in the
study of sigma models with $N=(2,2)$ supersymmetry 
\cite{Gates:1984nk,Howe:1985pm}, as these have  target spaces which  
necessarily  have a generalized \kah geometry. Such models have proven 
to be a powerful tool for elucidating further aspects of this geometry.
Away from certain loci (irregular points of certain canonical Poisson structures), 
generalized \kah geometry can be encoded (locally)
in terms of a single real function: the generalized \kah potential. In the 
language of supersymmetric sigma models, the generalized \kah potential is
the Lagrangian density in $N=(2,2)$ superspace 
\cite{Gates:1984nk,Buscher:1987uw,Sevrin:1996jr,Lindstrom:2005zr,Lindstrom:2007xv}.  
Being a potential, it is defined modulo certain 
ambiguities that can be understood both from the geometric 
and from the sigma model points of view. This paper is an 
attempt to understand the global issues related to the 
generalized \kah potential, and in particular the aspects that can be understood in terms of gerbes.

Gerbes are a geometrical realization of $H^3(M, \mathbb{Z})$ 
in a manner analogous to the way a line bundle is a geometrical 
realization of $H^2(M, \mathbb{Z})$. The notion of a holomorphic 
line bundle is closely related to K\"ahler geometry. In this paper, we define 
and investigate the properties of a structure that we call a {\em biholomorphic gerbe}.
A biholomorphic gerbe can be defined on a bicomplex manifold 
$(M, J_+, J_-)$, \ie a manifold $M$ equipped with two complex structures.
On such a manifold, a biholomorphic gerbe is a collection of
those transition functions defined on the triple intersections
\beq
G_{\alpha\beta\gamma}~:~U_\alpha\cap U_\beta 
\cap U_\gamma~\rightarrow~\mathbb{C}^*~,
\eeq{definitionbiholgerbe1}
which are biholomorphic\footnote{Our use of the word should 
not be confused with a bijective holomorphic function 
whose inverse is also holomorphic, which is sometimes also 
referred to as a biholomorphic function}, \ie holomorphic with 
respect to both complex structures. Moreover, the transition functions 
are antisymmetric under permutations of the open sets and satisfy 
a cocycle condition on fourfold intersections.  We give a precise 
definition of biholomorphic gerbes, and show that they arise naturally 
within generalized K\"ahler geometry. 
Where necessary, we shall assume that either $M$ is compact, or that
suitable boundary conditions are imposed.

Our analysis is motivated and guided by the sigma model discussion.  
One objective of the paper is to translate sigma model considerations 
into proper geometrical terms. 
To make the paper accessible to both physicists and mathematicians, we 
review some standard material
concerning  line bundles, gerbes and supersymmetric sigma models.  
In all our constructions we adopt the concrete and simple 
description of gerbes advocated by Hitchin  \cite{hitchinwhat, hitchingerbe}.  
Note that when we use the terms holomorphic (biholomorphic) 
functions and their exponentials, we have in mind $\cal{O}$, the sheaf of holomorphic 
(biholomorphic) functions, and $\cal{O}^*$, the multiplicative sheaf of nowhere zero 
holomorphic (biholomorphic) functions, respectively. 

Previous uses of gerbes in physics, particularly in the context of WZW models and anti-symmetric tensor gauge fields,
have appeared in, \eg
  \cite{Gawedzki:2002se,Gawedzki:2004tu,
Carey:2004xt, Aschieri:2004yz, Schreiber:2005mi,
Belov:2007qj}.

The paper is organized as follows. In Section \ref{realgerbe} 
we review some basic facts about line bundles and gerbes. 
Section \ref{holomorphicline} reviews holomorphic 
line bundles and their relation to \kah geometry.  
Section \ref{holomorphicgerbe} discusses the notion 
of a holomorphic gerbe with a Hermitian connection.  
We point out the appearance of a flat gerbe associated 
to a Hermitian connection.  In Section \ref{genkahler} we review 
the bihermitian description of generalized \kah geometry 
and discuss the properties of gerbes with connection associated to this 
geometry.  In particular, we show that generalized \kah geometry 
can be encoded in terms of two flat gerbes with additional 
very special properties.  Section \ref{potential} is a key 
part of the present work where we discuss the gluing of the 
generalized \kah potential and the relation to biholomorphic gerbes. 
Section \ref{BILP} deals with the special case in which the two 
complex structures commute; then all points are regular and 
the situation is particularly simple. Section \ref{summary} 
presents a summary of the results as well as some open questions. 
In the Appendix we discuss the example of the 
natural biholomorphic gerbe with connection on $S^3 \times S^1$. 

 \Section{Line bundles and gerbes}
\label{realgerbe}

In this section we review some standard facts about line bundles with 
connection and   gerbes with connection 
and introduce our notation.
We consider a smooth manifold $M$ with an open cover $\{ U_\alpha\}$ 
where all open sets and intersections are contractible.  

\subsection{Line bundles and $U(1)$ connections}
Let us first recall some facts about line bundles.
An  $S^1$-bundle can be thought of as a
set of transition functions
\beq
g_{\alpha\beta}~:~U_\alpha\cap U_\beta~\rightarrow~S^1~,
\eeq{definitiontransline}
which satisfy $g_{\alpha\beta} = g^{-1}_{\beta\alpha}$ 
and the cocycle condition on 
$U_\alpha\cap U_\beta \cap U_\gamma$
\beq
g_{\alpha\beta} g_{\beta\gamma} g_{\gamma\alpha} = 1~.
\eeq{cocycleline}
This condition is trivially satisfied when the transition 
functions are themselves   coboundaries: 
$g_{\alpha\beta}=h_\alpha h^{-1}_\beta$.
 $S^1$-bundles
are equivalent to complex line bundles with a Hermitian metric.

To any  $\frac{\omega}{2\pi} \in H^2(M, \mathbb{Z})$ 
we can associate a line bundle with connection as follows.
Using the Poincar\'e lemma, we find 
1-forms $A_\alpha$, functions $\Lambda_{\alpha\beta}$ and 
constants $ d_{\alpha\beta\gamma}$ satisfying 
\ber
\label{chainofrelations1line} 
&\omega= dA_\alpha~,~~~~~&A_\alpha \in \Omega^1(U_\alpha)~,\\
\label{chainofrelations2line} & A_\alpha - A_\beta = 
d \Lambda_{\alpha \beta}~,~~~~~&\Lambda_{\alpha\beta} 
\in C^\infty(U_\alpha\cap U_\beta)~,\\
\label{chainofrelations3line} & \Lambda_{\alpha\beta} 
+ \Lambda_{\beta\gamma} 
+ \Lambda_{\gamma\alpha} = d_{\alpha\beta\gamma}~,~~~~~& 
d_{\alpha\beta\gamma} \in 2\pi ~\mathbb{Z}
\eer{chainofrelationsline}
where the last relation is guaranteed since  $\frac{\omega}{2\pi} \in H^2(M, \mathbb{Z})$ 
(see \egg \cite{woodhouse} for the proof).  Since the coboundary of 
$\Lambda_{\alpha\beta}$ is equal to $2\pi$ times
an integer, we can exponentiate it to get   transition 
functions $g_{\alpha\beta} = e^{i\Lambda_{\alpha\beta}}$ 
that satisfy the cocycle condition (\ref{cocycleline}) 
on triple intersections.  The condition (\ref{chainofrelations2line})
can be rewritten as  
\beq
iA_\alpha - i A_\beta = g_{\alpha\beta}^{-1} d g_{\alpha\beta}
\eeq{connectionline}
and thus the set of one-forms $A_\alpha$ defines a connection on a line bundle and $\omega$ is its 
curvature.  The possible choices of inequivalent connection with the same curvature are 
parametrized by $H^1(M, \mathbb{R})/ H^1(M, \mathbb{Z})$. When the curvature vanishes, this is
precisely the space of flat connections, parameterized by their holonomies.

\subsection{Gerbes}
This definition of a line bundle and a connection can be generalized to gerbes. 
Gerbes were invented by Giraud \cite{giraud} and later 
extensively discussed by Brylinski  \cite{brylinski}.  We use the simple point of view
advocated by Hitchin in \cite{hitchingerbe}.   
Consider maps defined on each threefold intersection
\beq
g_{\alpha\beta\gamma}~:~ U_\alpha \cap U_\beta \cap 
U_\gamma ~\rightarrow ~S^1~,
\eeq{definitionmapgerbe}
satisfying 
\beq
g_{\alpha\beta\gamma} = g_{\beta\gamma\alpha} = g_{\gamma\alpha\beta}
= g^{-1}_{\beta\alpha\gamma} = g^{-1}_{\alpha\gamma\beta} =
g^{-1}_{\gamma\beta\alpha}
\eeq{antisymmetryfull}
as well as the cocycle condition on 
$U_\alpha\cap U_\beta\cap U_\gamma \cap U_\delta$
\beq
g_{\alpha\beta\gamma} g_{\beta\alpha\delta} 
g_{\gamma\beta\delta}g_{\delta\alpha\gamma}=1~.
\eeq{cocycleconditiondefin}
As for the line bundle, this condition is trivially satisfied when the transition functions are themselves coboundaries: 
$g_{\alpha\beta\gamma}=h_{\alpha\beta}h_{\beta\gamma}h_{\gamma\alpha}$.
This data defines a gerbe; we use this definition throughout, 
though there exist  other (equivalent) definitions; for details 
see \cite{hitchingerbe, chatterjee, grundle, murray}.

\subsection{$U(1)$ connections on gerbes}
By analogy with the line bundle case we can interpret
$\frac{H}{2\pi} \in H^3(M, \mathbb{Z})$
as the curvature\footnote{The requirement that the form be integral 
is an additional requirement on the geometry from the point of view of
mathematics. {}From the physics point of view this requirement 
is very natural since (on a compact target space) flux has to be
quantized to give a well defined quantum theory.} of a gerbe 
with connection. The Poincar\'e lemma implies the following
chain of relations
\ber
\label{chainofrelations1} &H= dB_\alpha~,~~~~~&B_\alpha \in \Omega^2(U_\alpha)~,\\
\label{chainofrelations2} & B_\alpha - B_\beta = 
d A_{\alpha \beta}~,~~~~~&A_{\alpha\beta} \in \Omega^1(U_\alpha\cap U_\beta)~,\\
\label{chainofrelations3} & A_{\alpha\beta} + A_{\beta\gamma} 
+ A_{\gamma\alpha} = d \Lambda_{\alpha\beta\gamma}~,~~~~~&
\Lambda_{\alpha\beta\gamma} \in C^\infty (U_\alpha\cap U_\beta \cap U_\gamma)~,\\
\label{chainofrelations4} &\Lambda_{\alpha\beta\gamma} + 
\Lambda_{\beta\alpha\delta} + \Lambda_{\gamma\beta\delta} +
\Lambda_{\delta\alpha\gamma}= d_{\alpha\beta\gamma\delta}~,~~~~~&
d_{\alpha\beta\gamma\delta} \in 2\pi ~\mathbb{Z}~,
\eer{chainofrelations}
where the last one is satisfied as a consequence of $\frac{H}{2\pi} 
\in H^3(M, \mathbb{Z})$.  Using this data we can define the set of functions 
$g_{\alpha\beta\gamma}~:~ U_\alpha \cap U_\beta \cap 
U_\gamma ~\rightarrow ~S^1$ given by
\beq
g_{\alpha\beta\gamma} = e^{i\Lambda_{\alpha\beta\gamma}}~,
\eeq{gerbetransitions}
which, as a result of   (\ref{chainofrelations4}), satisfy
(\ref{antisymmetryfull})-(\ref{cocycleconditiondefin}) and thus define a gerbe.
 Equation (\ref{chainofrelations3}) can be rewritten as
\beq
i A_{\alpha\beta} + i A_{\beta\gamma} + i A_{\gamma\alpha} 
= g^{-1}_{\alpha\beta\gamma} d g_{\alpha\beta\gamma}~.
\eeq{chainofrelations3extra}
{}From (\ref{gerbetransitions}) we see that $\Lambda_{\alpha\beta\gamma}$ are angles,
\begin{eqnarray}
\Lambda_{\alpha\beta\gamma} \in 2\pi {\mathbb R}/ {\mathbb Z}~.
\end{eqnarray}  
The above data defines (up to equivalence) a gerbe with connection.

\subsection{Transition line bundle description of gerbes}
\label{TransGerbe}
There is an alternative way to define gerbes 
\cite{chatterjee,hitchingerbe}. A gerbe can always be made trivial locally. 
What this means is that in each open
set $U_\alpha$ it is possible to choose an open cover 
$\{V_i^{(\alpha)}\}= U_\alpha\cap U_i$ and
functions $h^{(\alpha)}_{ij}$ defined on 
$V^{(\alpha)}_i\cap V^{(\alpha)}_j$ such that on
$V^{(\alpha)}_i\cap V^{(\alpha)}_j\cap V^{(\alpha)}_k$ we have
\beq
g_{ijk} = h^{(\alpha)}_{ij}h^{(\alpha)}_{jk}h^{(\alpha)}_{ki}.
\eeq{gijk1}
The choice of $h^{(\alpha)}_{ij}$ in general is different 
in each $U_\alpha$. Only when the gerbe is trivial can
one make such a choice globally. In the overlap $U_\alpha\cap U_\beta$ 
we now have two different trivializations
\beq
g_{ijk} = h^{(\alpha)}_{ij}h^{(\alpha)}_{jk}h^{(\alpha)}_{ki}
= h^{(\beta)}_{ij}h^{(\beta)}_{jk}h^{(\beta)}_{ki}
\eeq{gijk2}
Thus $f_{ij} = h^{(\alpha)}_{ij}/h^{(\beta)}_{ij}$
satisfies the cocycle condition which implies that $f_{ij}$ are the transition
functions of a line bundle defined on $U_\alpha\cap U_\beta$. This line bundle 
is called the transition line bundle of the gerbe, and is an equivalent way 
of encoding the data of the gerbe.

\subsection{Flat gerbes}
A {\em flat} gerbe is defined as a gerbe with vanishing curvature: $H=0$. 
Then $B_\alpha$ is closed, so that using the Poincar\' e lemma, 
\ber
\label{arealgerbflat1}   B_\alpha &=& d q_\alpha   ~,\\
\label{arealgerbflat2}    A _{\alpha\beta} &=& q_\alpha - 
q_\beta + d p_{\alpha\beta}  ~,\\
\label{arealgerbflat3}   \Lambda _{\alpha\beta\gamma} &=& p_{\alpha\beta} +
p_{\beta\gamma} + p_{\gamma\alpha}+ l_{\alpha\beta\gamma}~,\\
\label{achainofrelations4} l_{\alpha\beta\gamma} + l_{\beta\alpha\delta} + 
l_{\gamma\beta\delta} + l_{\delta\alpha\gamma} &=& 
d_{\alpha\beta\gamma\delta}~,~~~~~d_{\alpha\beta\gamma\delta}
\in 2\pi ~\mathbb{Z}~,
\eer{flatgerbey}
for some $q_\alpha$ which  are one-forms on $U_\alpha$, some $p_{\alpha\beta}$ which  are functions 
on $U_\alpha \cap U_\beta$  and some  $l_{\alpha\beta\gamma}$ which are 
constants.  Since the $  \Lambda _{\alpha\beta\gamma}$ are angles,
the constants $l_{\alpha\beta\delta} $ are also angles, only determined up to an 
additive factor $2\pi\mathbb{Z}$:
\begin{eqnarray}
l_{\alpha\beta\gamma} \in 2\pi {\mathbb R}/ {\mathbb Z}~.
\end{eqnarray}  
Then $\exp {(il_{\alpha\beta\gamma})} $ is a 
 2-cocycle  so that $l_{\alpha\beta\gamma}/2\pi$ is a 2-cocycle in $ {\mathbb R}/{\mathbb Z}$, and
 represents a \v{C}ech 
class in $H^2(M, {\mathbb R}/{\mathbb Z})$, which corresponds
to the holonomy of the flat gerbe.   
A flat gerbe is then defined by   transition functions of the form
\beq
g_{\alpha\beta\gamma} = e^{i(p_{\alpha\beta} +
p_{\beta\gamma} + p_{\gamma\alpha})}e^{i l_{\alpha\beta\gamma}}
\eeq{flatgerbetransitions}
which is the product of a trivial piece and a constant.

\Section{Holomorphic line bundles}
\label{holomorphicline}
Complex holomorphic line bundles over complex manifolds are natural structures
in the study of K\"ahler geometry.
\subsection{Holomorphic line bundles and Hermitian connections}
When $M$ is a complex manifold, 
a {\em holomorphic} line bundle can be defined as a set of holomorphic functions
\beq
G_{\alpha\beta}~:~U_\alpha\cap U_\beta~\rightarrow~\mathbb{C}^*~,
\eeq{definhollinebasic}
with $G_{\beta\alpha}\equiv(G_{\alpha\beta})^{-1}$ obeying a 
cocycle condition  of the form (\ref{cocycleline}) on triple intersections. The topology of the 
bundle is classified by $H^2(M, \mathbb{Z})$, and a class is represented by
a two-form $\frac{\omega}{2\pi} \in H^2(M, \mathbb{Z})$. One can then define 
an underlying line bundle with connection whose curvature is $\omega$. 
If furthermore $\omega$ is of type $(1,1)$
with respect to the complex structure, (locally) we can write 
\beq
\omega = i \partial \bar{\partial} K_\alpha = \frac{1}{2} d d^c K_\alpha~,
\eeq{formKahpoet}
where $K_\alpha$ is a real function on $U_\alpha$, 
defined up to shifts by the real part of a holomorphic function,
and $d^c\equiv i(\bar{\partial}-\partial)$.  On $U_\alpha \cap U_\beta$ we 
have 
\beq
K_\alpha - K_\beta = F_{\alpha\beta} (z) + \bar{F}_{\alpha\beta}(\bar{z})~, 
\eeq{doubleline}
where $F_{\alpha\beta}$ is a holomorphic function on $U_\alpha \cap U_\beta$.
On the triple intersection $U_\alpha \cap U_\beta \cap U_\gamma$, 
eq.~(\ref{doubleline}) implies
\beq             
Re \left(F_{\alpha\beta} + F_{\beta\gamma} + F_{\gamma\alpha}\right)=0~.
\eeq{relationrealline}
Comparing these relations with the real equations
(\ref{chainofrelations1line})-(\ref{chainofrelations3line}), we find
\beq
A_\alpha = \frac{1}{2} d^c K_\alpha~,~~~~~~~~
\Lambda_{\alpha\beta} = Im(F_{\alpha\beta})~.
\eeq{realrelationcile}
Thus, on triple intersections, the holomorphic function $F_{\alpha\beta}$ satisfies
\beq
F_{\alpha\beta} + F_{\beta\gamma} + F_{\gamma\alpha} = i d_{\alpha\beta \gamma} \in
2\pi i \mathbb{Z}~,
\eeq{indetityholline}
which allows us to define the holomorphic transition functions 
\beq
G_{\alpha\beta} (z) = e^{F_{\alpha\beta}(z)}~:~U_\alpha \cap 
U_\beta~\rightarrow~ \mathbb{C}^*~.
\eeq{fgtrans}
These transition functions satisfy the standard cocycle condition  of the form (\ref{cocycleline})
and thus define a holomorphic line bundle.  Furthermore, $e^K$ has transition functions
$
 e^{K_\alpha} =G_{\alpha\beta}  \bar{G}_{\alpha\beta} e^{K_\beta}
$
which are precisely the transition functions that a   Hermitian fiber metric should have.
Thus $e^K$ is a Hermitian fiber metric and so defines a Hermitian structure on the holomorphic line bundle; 
such a structure exists whenever
$\frac{\omega}{2\pi} \in H^2(M, \mathbb{Z})$ 
is of type $(1,1)$. The condition satisfied by the transition functions can be rewritten as
\beq
G_{\alpha\beta}  \bar{G}_{\alpha\beta} = e^{K_\alpha} e^{-K_\beta}~.
\eeq{hermitianlinebundle}
with the right hand side a trivial cocycle, and this is the form of the condition we generalize to gerbes. 

\subsection{Relation to K\"ahler-Hodge geometry}
Holomorphic line bundles play an important role in K\"ahler-Hodge geometry. 
Consider the \kah manifold $(M, J, g)$ with $\omega=gJ$ being a \kah form.
Then formula (\ref{formKahpoet}) provides the local definition of the \kah potential. 
The global information about the geometry is encoded in an underlying 
holomorphic line bundle equipped with a Hermitian metric. 
When $\omega/2\pi \in H^2(M, \mathbb{Z})$ the manifold is said to be Hodge
and the \kah potential can be given  as  
\beq
K_\alpha =
\log || s_\alpha ||^2~,
\eeq{definitionofkahpot}
where $s_\alpha$ is a nowhere vanishing section of a 
holomorphic line bundle and $||s||^2 = h s \bar{s}$ with 
$h$ the Hermitian metric on this bundle \cite{Hitchin:1986ea}.   

\Section{Holomorphic gerbes and connections}
\label{holomorphicgerbe}
For complex manifolds, one can define holomorphic gerbes in complete
analogy with holomorphic line bundles.

\subsection{Holomorphic gerbes}
A {\it holomorphic} gerbe on a complex manifold $M$ is a set of holomorphic functions
\beq
G_{\alpha\beta\gamma}~:~U_\alpha\cap U_\beta 
\cap U_\gamma~\rightarrow~\mathbb{C}^*
\eeq{definitionholgerbe}
that are antisymmetric under permutations of the open sets and satisfy 
a cocycle condition on fourfold intersections.  Moreover, if there exist 
real functions  $h_{\alpha\beta}$ on double intersections such that 
\beq
G_{\alpha\beta\gamma} \bar{G}_{\alpha\beta\gamma} = 
h_{\alpha\beta} h_{\beta\gamma} h_{\gamma\alpha}
\eeq{definitionhermitiangerbe}
so that $G \bar{G}$ is a trivial cocycle, then we refer to such a  gerbe as  a holomorphic gerbe with a Hermitian structure, as this is a natural generalisation of the condition (\ref{hermitianlinebundle}) for hermitian structures on holomorphic  line bundles.


\subsection{Hermitian connections on holomorphic gerbes}
For a closed 3-form $H$ such that $\frac{H}{2\pi} \in 
H^3(M, \mathbb{Z})$ there exists a gerbe
with connection as described in Section \ref{realgerbe}.
Assume that $H$ is of type  $(2,1)+(1,2)$ with respect to the 
complex structure. Below we explain that 
this  gives a generalization of    the holomorphic line bundle 
which
corresponds to a holomorphic gerbe with a Hermitian 
structure and a connection that respects the Hermitian 
structure; we refer to this as a holomorphic gerbe with
hermitian connection. 

On $U_\alpha$, a connection two-form 
$B_{\alpha}$ with $H=dB_{\alpha}$ can be chosen to be of type $(1,1)$
(we refer to this as the $(1,1)$ gauge for $B$). 
Then on  the double intersection 
$U_\alpha \cap U_\beta$ we have that $B_{\alpha}^{(1,1)}  - B_{\beta}^{(1,1)}$ is closed 
and $d^c$-closed so that it can be written as  
$i \bar{\partial}
\partial \upsilon_{\alpha\beta}  $ for some real function  $ \upsilon_{\alpha\beta}$
on $U_\alpha \cap U_\beta$.
For later convenience, we   write this as the real part of some complex function
$\xi_{\alpha\beta}$ on $U_\alpha \cap U_\beta$,  
$\upsilon_{\alpha\beta}=\xi_{\alpha\beta} + \bar{\xi}_{\alpha\beta}$, since the imaginary part   plays  a role later.
Then
\beq
B_{\alpha}^{(1,1)}  - B_{\beta}^{(1,1)} = i \bar{\partial}
\partial (\xi_{\alpha\beta} + \bar{\xi}_{\alpha\beta}) ~,
\eeq{coB11form}
and on 
the triple intersection $U_\alpha \cap U_\beta \cap U_\gamma$ we find
\beq
\xi_{\alpha\beta} + \xi_{\beta\gamma} + \xi_{\gamma\alpha} 
+ \bar{\xi}_{\alpha\beta} + \bar{\xi}_{\beta\gamma} + \bar{\xi}_{\gamma\alpha}= 
- f_{\alpha\beta\gamma}(z) - \bar{f}_{\alpha\beta\gamma}(\bar{z})~,
\eeq{complexxiform}
with $f_{\alpha\beta\gamma}$  a holomorphic function on 
$U_\alpha\cap U_\beta \cap U_\gamma$. Comparing   with (\ref{chainofrelations1})-(\ref{chainofrelations3}) 
we find the following relations
\ber
\label{realgerbhol2}    A_{\alpha\beta} &= & \frac{i}{2}
\partial (\xi_{\alpha\beta} + \bar{\xi}_{\alpha\beta}) 
-\frac{i}{2} \bar\partial (\xi_{\alpha\beta} + \bar{\xi}_{\alpha\beta})~,\\
\label{realgerbhol3}   \Lambda_{\alpha\beta\gamma}& = &\frac{i}2 
( \bar{f}_{\alpha\beta\gamma} - f_{\alpha\beta\gamma}) ~.
\eer{definitionLambdaholgerbe}

In analogy with the holomorphic line bundle, the 
imaginary part of the holomorphic 
function $f$ satisfies (\ref{chainofrelations4})
whereas the real part is a trivial cocycle as a consequence of (\ref{complexxiform}). Therefore 
\beq
f_{\alpha\beta\gamma} + f_{\beta\alpha\delta} + f_{\gamma\beta\delta} +
f_{\delta\alpha\gamma} \in 2\pi i ~\mathbb{Z}~, 
\eeq{imaginaryholgerbeco}
and we can define holomorphic transition functions
\beq
G_{\alpha\beta\gamma} = e^{f_{\alpha\beta\gamma}}
\eeq{holomogerbetransf}
that satisfy the cocycle condition of the form (\ref{cocycleconditiondefin}) on fourfold intersections. Moreover, due to the 
property (\ref{complexxiform}), the corresponding 
holomorphic gerbe has a Hermitian structure, \ie the transition functions 
satisfy (\ref{definitionhermitiangerbe}) 
with $h_{\alpha\beta} = \exp (- \xi_{\alpha\beta} - \bar{\xi}_{\alpha\beta})$. 
This is then a holomorphic gerbe with Hermitian connection.
The curvature three-form $H$ is necessarily of type $(2,1)+(1,2)$ with respect to 
the complex structure.

We are  particularly interested in   generalized  (K\"ahler) geometries for which the closed form $H$ also  satisfies $d^c H=0$, so that $\partial H =\bar{\partial} 
H=0$.
Locally the $(2,1)$-part of such an $H$ can be written as follows
\beq
H^{(2,1)} = i \partial \bar{\partial} \lambda^{(1,0)}_\alpha~,
\eeq{21partofHholgerbe}
where $\lambda^{(1,0)}_\alpha$ is a complex $(1,0)$-form on $U_\alpha$.  
Alternatively, in real coordinates we can write 
\beq
H = d d^c (Re~ \lambda^{(1,0)}_\alpha)~.
\eeq{realHholgerbe}
Choosing  the $(1,1)$ gauge for $B$ and using
(\ref{21partofHholgerbe}), we obtain
\beq
B_{\alpha}^{(1,1)} = i \bar{\partial} \lambda_{\alpha}^{(1,0)} - 
i \partial \bar{\lambda}_\alpha^{(0,1)}~.
\eeq{realgerbhol1}
On the double intersection $U_\alpha \cap U_\beta$ we have, using (\ref{coB11form}),
\beq
\lambda^{(1,0)}_\alpha - \lambda^{(1,0)}_\beta =  \partial \xi_{\alpha\beta} 
+ \phi_{\alpha\beta}^{(1,0)}~,
\eeq{doubleholgerbe}
where $\xi_{\alpha\beta}$ is a complex function whose real part enters  in (\ref{coB11form}) and $\phi^{(1,0)}$ 
is a holomorphic $(1,0)$-form (\ie $\bar{\partial} \phi_{\alpha\beta}^{(1,0)}=0$). 
This decomposition is not unique: we can always shift a $\partial$ exact 
holomorphic one-form between the two terms on the RHS of (\ref{doubleholgerbe}), while leaving (\ref{coB11form}) unchanged.

\subsection{Transition line bundle on a holomorphic gerbe}
Notice that from (\ref{coB11form}) we can interpret $-(\xi + \bar{\xi})$ 
as the (K\"ahler) potential
for the curvature $\delta B_{\alpha\beta} = B_\alpha - B_\beta$ of the 
transition line bundle defined on $U_\alpha\cap U_\beta$.
This gives an equivalent description of a holomorphic gerbe
with Hermitian connection in terms of a holomorphic transition line bundle with 
hermitian structure ({\it cf.} the discussion 
in section 
\ref{TransGerbe}
  for the
real case).

\subsection{Hermitian geometry of $(2,0)$-supersymmetric sigma models}
The Hermitian connection on the holomorphic gerbe described above
defines a Hermitian geometry $(M,J,g)$ with 
complex structure $J$ and Hermitian metric $g$, which is 
precisely the geometry of a sigma model  with (2,0) supersymmetry 
\cite{Hull:1985jv}. The fundamental 2-form $\omega=gJ$ is 
of type $(1,1)$ but is not closed. Instead, it satisfies
\beq
dd^c \omega =0~,
\eeq{fun}
and defines a torsion 3-form
\beq
H=d^c\omega~,
\eeq{sdfsd}
which is closed and $d^c$-closed and of type $(2,1)+(1,2)$, so that it is given in terms of a 1-form potential  $ {\lambda}^{(1,0)}_\alpha$ by
(\ref{realHholgerbe}).
The fundamental two-form is given in $U_\alpha $ in terms of   $ {\lambda}^{(1,0)}_\alpha$ by
\beq
\omega = -( \bar{\partial} {\lambda}^{(1,0)}_\alpha + 
\partial \bar\lambda^{(0,1)}_\alpha)~.
\eeq{lrtio}
If $ {\lambda}^{(1,0)}_\alpha = - i \partial K_\alpha$, then the 
manifold is K\"ahler with K\"ahler potential
$K_\alpha$ and $H=0$.

\subsection{An associated flat gerbe}
Because the gerbe curvature (\ref{realHholgerbe}) 
has 1-form potential $Re (\lambda_{\alpha}^{(1,0)})$,
we can define a collection of locally defined closed forms
\beq
{\cal F}_\alpha = d ~Re (\lambda_{\alpha}^{(1,0)})~.
\eeq{definflatgerbe}
We   interpret this collection of two-forms as connection 
two-forms for a flat gerbe.  We elaborate on this flat gerbe in the following sections.

 \Section{Generalized \kah geometry}
\label{genkahler}  

In this Section we review the definition of generalized \kah 
geometry and discuss the gerbe associated with this geometry. 

\subsection{Bihermitian formulation of generalized \kah geometry}
A generalized \kah manifold $(M, J_+, J_-, g)$ is a manifold $M$ with
two complex structures $J_\pm$ and a bihermitian metric $g$ satisfying the 
integrability conditions
\beq
d_+^c \omega_+ + d_-^c \omega_-=0~,~~~~~
d d_\pm^c \omega_\pm = 0~,
\eeq{integrabilityforGKS} 
where $\omega_\pm =gJ_\pm$ and $d_\pm^c$ are the $i(\bar{\partial}-\partial)$ operators associated 
with the complex structures $J_\pm$.  The conditions (\ref{integrabilityforGKS}) imply
that we can define a closed three form
\beq
H = d_+^c \omega_+ = - d_-^c \omega_-~,~~~ dH = 0
\eeq{definitionofH}
which is also $d^c$-closed, $d^cH=0$.

\subsection{Gerbes on a generalized \kah geometry}
If $\frac{H}{2\pi} \in H^3(M, \mathbb{Z})$ then we have a gerbe with connection.
The condition $\frac{H}{2\pi} \in H^3(M, \mathbb{Z})$  is necessary for 
the sigma model with Wess-Zumino term specified by $H$ to give a well-defined 
quantum theory (on a compact target space), and we assume that this holds.
The definitions $\omega_\pm =gJ_\pm$ and  (\ref{definitionofH}) imply that
$H$ is a $(2,1)+(1,2)$ form with respect to {\em both} complex structures:
\ber
H = H^{(2,1)}_+ +H^{(1,2)}_+ = H^{(2,1)}_- +H^{(1,2)}_-~,\\
H^{(2,1)}_\pm  - H^{(1,2)}_\pm = \mp i d\omega_\pm~,
\eer{definition21partHKah}
which implies 
\beq
H_+^{(2,1)} - H_-^{(2,1)} = H_-^{(1,2)} - H_+^{(1,2)}
= -\frac{i}{2} d \left ( \omega_+ + \omega_- \right )~,
\eeq{sskw200120}
\beq
H_+^{(2,1)} - H_-^{(1,2)} = H_-^{(2,1)} - H_+^{(1,2)}
= -\frac{i}{2}  d \left (\omega_+ - \omega_- \right )~.
\eeq{HHHhol33992}

\subsection{Two gerbe connections}
As $H$ is a $(2,1)+(1,2)$ form with respect to both complex 
structures, the discussion from the previous section about 
holomorphic hermitian gerbes applies twofold: the globally defined 
curvature 3-form $H$ is the curvature of two holomorphic gerbes with 
Hermitian connection, one associated with each of the two complex 
structures, $J_\pm$. We use the same notation as in Section \ref{holomorphicgerbe}, 
adding the subscript $\pm$ to indicate the relevant complex structure.  For a 
given $H$ we can choose a connection $B_{+\alpha} $ that is $(1,1) $ respect to $J_+$
 or a connection $B_{-\alpha}$
which is $(1,1) $ respect to $J_-$, with   $H=d B_{\pm\alpha}$.
 Note that in general these   specify 
inequivalent gerbes with connection.

For each choice of connection $B_\pm$ there are descendants 
$\lambda_{\pm}^{(1,0)}, \xi_\pm, f_\pm$
satisfying formulas of the form (\ref{realgerbhol1})-(\ref{realgerbhol3}) with $\pm$ 
added appropriately to indicate the choice of complex structure.


\subsection{A flat gerbe}
\label{sectionflatgerbe}
Two gerbes with connection associated to the same curvature three-form 
differ by a flat gerbe, 
so $H=dB^{(1,1)}_{+\alpha} =dB^{(1,1)}_{-\alpha} $ and hence
$d[B^{(1,1)}_{+\alpha}- B^{(1,1)}_{-\alpha}]=0$. 
Then we have ({\it cf.} 
\ref{arealgerbflat1}-\ref{arealgerbflat3})
\ber
\label{realgerbflat1}   B^{(1,1)}_{+\alpha} - B^{(1,1)}_{-\alpha} = d q_\alpha   ~,\\
\label{realgerbflat2}    A^+_{\alpha\beta} - A^-_{\alpha\beta}= 
q_\alpha - q_\beta + d p_{\alpha\beta}  ~,\\
\label{realgerbflat3}   \Lambda^+_{\alpha\beta\gamma} -   
\Lambda^-_{\alpha\beta\gamma} = p_{\alpha\beta} + p_{\beta\gamma} + p_{\gamma\alpha}
+ l_{\alpha\beta\gamma}~,
\eer{flathermih}
where $q_\alpha$ are one-forms on $U_\alpha$, $p_{\alpha\beta}$ 
are functions on $U_\alpha \cap U_\beta$  and $l_{\alpha\beta\gamma}$
are constants.   
Using (\ref{complexxiform}) and the definition (\ref{holomogerbetransf}) 
we conclude that  the transition functions 
\begin{eqnarray}
g_{\alpha\beta\gamma}=G^+_{\alpha\beta\gamma} \bar{G}^-_{\alpha\beta\gamma}=
e^{i( \Lambda^+_{\alpha\beta\gamma} -   
\Lambda^-_{\alpha\beta\gamma})}
\end{eqnarray}
correspond to a flat gerbe, as they are of the form    (\ref{flatgerbetransitions}).  

However, generalized \kah geometry contains more structure than 
two Hermitian gerbes with Hermitian connections that have the same 
curvature. In particular, (\ref{definitionofH}) implies $H$ is $d^c_\pm$-exact.
The $\omega_\pm$  are $(1,1)$ forms  and they have nice expressions in 
terms of one-form potentials $\lambda_\pm^{(1,0)}$ (see the previous section)
\beq
\omega^{(1,1)}_{\pm} = 
\mp\left(\partial_\pm \bar\lambda_{\pm}^{(0,1)} +
\bar\partial_\pm\lambda_{\pm}^{(1,0)}\right)~.
\eeq{omegalambda}
Since $\omega_\pm$ are globally defined two forms  
we can conclude from (\ref{doubleholgerbe}) and (\ref{omegalambda}) that in  $U_\alpha \cap U_\beta$ 
$$ d d_\pm^c \left(\bar\xi_\pm-\xi_\pm \right) = 0~,$$
which tells us that the imaginary part of $(\xi_{\alpha\beta})_\pm$ 
can be written as the imaginary part of a holomorphic function for both complex structures. 
This in turn tells us that $\partial_\pm Im \xi_\pm$ is a holomorphic one-form and
can therefore be absorbed in $\phi_\pm$ in equation (\ref{doubleholgerbe}). 
Thus we conclude that $\xi$ can always be chosen to be real.
With a real $\xi$, (\ref{doubleholgerbe}) and (\ref{complexxiform}) imply that
\beq
 \phi^{(1,0)}_{\alpha\beta} + \phi ^{(1,0)}_{\beta\gamma} + \phi ^{(1,0)}_{\gamma\alpha}
=\frac{1}{2} \partial f_{\alpha\beta\gamma}=
 \frac{1}{2} G_{\alpha\beta\gamma}^{-1} d G_{\alpha\beta\gamma}~, 
\eeq{lksjkfhkhsk}

It is possible to choose a 
 two-form connection $B$ of type $(2,0)+(0,2)$ so that 
$H^{(2,1)} = dB^{(2,0)}_\alpha$. \footnote{We refer to this as 
the $(2,0)$ gauge. See, \eg \cite{Hull:2008vs} for this and other gauges.} 
The reality of $\xi$ allows us to choose an $A^{(1,0)}_{\alpha\beta}$ with 
the following chain of relations
\ber
\label{chainofrelationshol1} &H^{(2,1)} = dB^{(2,0)}_\alpha~,  \\
\label{chainofrelationshol2} & B^{(2,0)}_\alpha - B^{(2,0)}_\beta = d A^{(1,0)}_{\alpha \beta}~, \\
\label{chainofrelationshol3} & A^{(1,0)}_{\alpha\beta} + A^{(1,0)}_{\beta\gamma} + A^{(1,0)}_{\gamma\alpha}
= - \frac{i}{2} G_{\alpha\beta\gamma}^{-1} d G_{\alpha\beta\gamma}~, 
\eer{chainofrelationsholom}
where we have suppressed the labels $\pm$ denoting the complex structure. 
We stress that these relations do {\em not} hold in general 
for holomorphic gerbes with Hermitian connection. 

These relations imply that 
\beq
\partial B^{(2,0)}_\alpha=0, \qquad \bar \partial A^{(1,0)}_{\alpha\beta}=0~.
\eeq{dfsoppoipods}
From (\ref{21partofHholgerbe}), we see that $B^{(2,0)}_\alpha$ can be chosen to be
\beq
B^{(2,0)}_\alpha= -i\partial   \lambda^{(1,0)}_\alpha~
\eeq{btwois}
so that (\ref{doubleholgerbe}) implies that a holomprphic $(1,0)$ form satisfying (\ref{chainofrelationshol2}) is
\beq
A^{(1,0)}_{\alpha\beta}= 
-i\phi_{\alpha\beta}^{(1,0)}~.
\eeq{doubleholgerbewerwrt}
Then (\ref{chainofrelationshol3}) follows from (\ref{lksjkfhkhsk}).


\subsection{Two locally defined symplectic forms}
Combining the relations (\ref{definitionofH}) and (\ref{chainofrelationshol1}) 
we are led to a very unexpected description of generalized \kah geometry 
in terms of locally defined closed nondegenerate forms, which can also be
thought of in a more global language as flat gerbe connections.
Using the relations 
(\ref{lrtio}) and (\ref{btwois}), the  two-forms (\ref{definflatgerbe})  
can be written as   
\beq
{\cal F}^\pm_\alpha  =\frac{i}{2} \left ( B_{\pm\alpha}^{(2,0)} - 
B_{\pm\alpha}^{(0,2)} \right ) \mp \frac{1}{2} \omega_\pm~. 
\eeq{flatgerbeanother}
This is a collection of closed two-forms  ${\cal F}^\pm_\alpha$ 
defined on $U_\alpha$; $d{\cal F}^\pm_\alpha=0$
follows immediately from (\ref{definitionofH}) and (\ref{chainofrelationshol1}). 
Moreover, these forms are nondegenerate, \ie ${\cal F}^\pm_\alpha$ are symplectic 
structures on $U_\alpha$.  These forms (and linear combinations of them)
can be interpreted as connection forms for a flat gerbe, 
which may or may not be trivial. It
has been shown that the flat gerbes specified by the collection of two forms
$$ \frac{1}{2} \left ( {\cal F}^+_\alpha  \pm   {\cal F}^-_\alpha  \right ),$$
plays an essential role in the definition of topological
string theory on generalized K\"ahler manifolds \cite{Hull:2008vs}. 
The different choices of sign correspond to either A- or B-twist topological models.

\subsection{Another characterization of generalized \kah geometry}
The forms ${\cal F}^\pm_\alpha$ encode the full local geometrical 
data of the generalized \kah geometry on a bicomplex manifold. 
We have the following theorem holding locally:
\begin{theorem}
Consider a coordinate patch  $U$ of a bicomplex manifold.
Suppose there exist two symplectic forms ${\cal F}^\pm$
that tame the complex structures $J_\pm$ respectively, 
\ie for any nonzero tangent vector $v$ we have 
$${\cal F}^\pm (v,  J_\pm v) > 0~.$$
If in addition
$$ {\cal F}^+ J_+ - J_-^t {\cal F}^- ~$$
is a closed two-form, then this data defines a generalized \kah geometry on $U$.
\end{theorem}
{\it Proof:~} The proof is straightforward. Note that the condition in the theorem is
that
${\cal F}^+ (v,  J_+ v) > 0$ and ${\cal F}^- (v,  J_- v) > 0$.
Let us decompose the 2-form ${\cal  F}^+$ with respect to the complex structures $J_+$
and 
${\cal  F}^-$ with respect to $J_-$,
\beq
{\cal F}^\pm = ({\cal F}^\pm)^{(2,0)} + ({\cal F}^\pm)^{(0,2)} +  ({\cal F}^\pm)^{(1,1)}~.
\eeq{decompskkwo2}
Comparing with (\ref{flatgerbeanother}), we identify 
\beq
({\cal F}^\pm)^{(1,1)} = \mp \frac{1}{2} \omega_\pm~,~~~~~~~
({\cal F}^\pm)^{(2,0)}  = \frac{i}{2} (B^\pm)^{(2,0)}~. 
\eeq{identificationgb} 
Using these identifications we decompose
 ${\cal F}^+ J_+ $ and $J_-^t {\cal F}^-$ into symmetric and anti-symmetric parts,
\beq
{\cal F}^+ J_+ = - \frac{1}{2} B^+ + \frac{1}{2} g_+~,~~~~~~~
J_-^t {\cal F}^- = - \frac{1}{2} B^- + \frac{1}{2} g_-~,
\eeq{expdieoeoeo}
where $g_\pm$ are symmetric tensors and $B^\pm$ are anti-symmetric tensors. If
the corresponding symplectic structures tame the complex structures $J_\pm$, then
$g_\pm$ are positive-definite and so define metrics. The second condition in the theorem implies that 
\beq
g_+ = g_- \equiv g~,~~~~~~~~ dB^+ = d B^-\equiv H~.
\eeq{definsosoppwq}
Then  the condition $d{\cal F}^\pm=0$ implies 
\beq
dB^+  = d_+^c \omega_+ = dB^-= - d_-^c \omega_-~,
\eeq{finallly2999}
with $\omega_\pm = gJ_\pm$. Thus we obtain the standard local
bihermitian formulation of generalized \kah geometry.

This theorem can be understood as an alternative local description of   
generalized \kah geometry; it is formulated entirely in terms of locally 
{\em bisymplectic} bicomplex geometry. In order to make the theorem global
one has to specify the way the forms ${\cal F}^\pm$ patch together in overlaps.
As should be clear from the previous discussion, they have the transition functions of
flat gerbe connections.
A special case of this theorem
in which ${\cal F}^\pm$ are globally defined and obey ${\cal F}^+ = - {\cal F}^-$
was considered in \cite{poisson-gualtieri}.

In this Section we have presented a number of  results concerning
the gerbe structures of generalized \kah geometry.  However using 
the notion of the generalized \kah potential we can analyse the underlying 
structures further.

 \Section{The generalized \kah potential}
\label{potential}

In \cite{Gates:1984nk}, it was found that the $N=(2,2)$ superspace 
Lagrangian of supersymmetric sigma models encoded particular examples
of generalized \kah geometry in terms of a single generalized \kah potential; this was
extended to more generic situations in \cite{Buscher:1987uw}, and conjectured
to hold generally in \cite{Sevrin:1996jr}. 
In \cite{Lindstrom:2005zr}, it was proved that this is indeed the case: away from 
irregular points, a generalized \kah manifold can be locally described in terms of
a single real function $K$,  the generalized \kah potential. In this Section we discuss
the gluing properties of the generalized \kah potential and their relation to the underlying 
gerbe with connection. 

\subsection{Review of the potential}
We now briefly review the local geometry of a generalized \kah manifold and its description 
in terms of a potential.  Generalized \kah manifolds have a rich underlying Poisson geometry
\cite{Lyakhovich:2002kc, Hitchin:2005cv}: there are two real Poisson structures $\pi_\pm 
= (J_+ \pm J_-)g^{-1}$. We call a point regular if there exists a neighborhood of that point where 
$\pi_\pm$ have constant rank.  In addition there is a third Poisson
structure $\sigma = [J_+, J_-]g^{-1} = \pi_- g \pi_+$ which can be written as the
real part of a Poisson structure that is holomorphic with respect to either complex 
structure: $\sigma=\frac12[\sigma_\pm^{(2,0)}+\sigma_\pm^{(0,2)}]$.
In the neighborhood of a regular point we can introduce   coordinates adapted 
to the corresponding symplectic foliations: we introduce coordinates 
$(\phi, \bar{\phi}, \chi, \bar{\chi}, X_L, \bar{X}_L, X_R, \bar{X}_R)$ such
that $(d\phi, d\bar{\phi})$ span the kernel of $\pi_-$ and $(d\chi, d\bar{\chi})$ 
span the kernel of $\pi_+$. The remaining coordinates 
$X_{L,R}, \bar{X}_{L,R}$ lie along the leaves of $\sigma$. 
We choose Darboux coordinates $X_L,Y_L$ for 
$\sigma_+^{(2,0)}=dX_L\wedge d Y_L$ and $X_R,Y_R$ for 
$\sigma_-^{(2,0)}=dX_R\wedge dY_R$; then a polarization is
just a choice of an equal number of $X_L$ and $X_R$ coordinates out of the 
set $\{X_L,Y_L,X_R,Y_R\}$. 
The generalized K\"ahler potential is locally a function of 
$(\phi, \bar{\phi}, \chi, \bar{\chi}, X_L, \bar{X}_L, X_R, \bar{X}_R)$, and 
all geometrical quantities are given in terms of second derivatives of $K$. 
The relations are linear if $\sigma=0$ (see the next section) but are nonlinear in general.
For further details the reader can consult 
\cite{Lindstrom:2005zr, Lindstrom:2007qf, Lindstrom:2007xv}.

The generalized \kah potential $K$ is not defined uniquely by the geometry;
the precise form of the relation between  $K$ and the geometry depends on the choice of polarization on the leaves of $\sigma$, and for a given polarization
$K$ can be shifted 
 by generalized \kah gauge transformations 
 without changing the geometry, as is
  described below. 
  A change of a polarization on 
a leaf of $\sigma$ corresponds to a coordinate change (symplectomorphism)
and transforms $K$ by a Legendre transformation. 

We now focus on generalized \kah manifolds that are 
regular everywhere; the case with irregular points is more subtle and we leave
it to future investigations. 

\subsection{Generalized \kah transformations on overlaps}
When all the Poisson structures are regular everywhere
on the manifold, we can choose natural coordinates in each patch and in their 
intersections.  In addition, $K$ can transform when we move
between different patches. For instance, 
if we have the same polarizatrion in $U_\alpha $ and $ U_\beta$, then 
requiring that 
$K_\alpha $ and $ K_\beta$ define the same geometry on $U_\alpha \cap U_\beta$ 
 implies
\beq
K_\alpha - K_\beta = F^+_{\alpha\beta} (\phi, \chi, X_L) +
\bar{F}^+_{\alpha\beta} (\bar{\phi}, \bar{\chi}, \bar{X}_L) 
+ F^-_{\alpha\beta}(\phi, \bar{\chi}, X_R) + 
\bar{F}^-_{\alpha\beta} (\bar{\phi}, \chi, \bar{X}_R)~,
\eeq{doubleintersection}
for some  special
  $J_+$-holomorphic function $F^+_{\alpha\beta}$
and special   $J_-$-holomorphic function $F^-_{\alpha\beta}$.\footnote{An, \eg $J_+$-holomorphic function
could depend on $(\phi, \chi, X_L, Y_L)$; a {\it special} $J_+$-holomorphic function
is defined with respect to a choice of polarization and depends only on $(\phi, \chi, X_L)$.}
These functions 
are defined in turn up to the following shifts
\ber
\label{shiftambiguity1}  F^+_{\alpha\beta} (\phi, \chi, X_L) &\rightarrow &
F^+_{\alpha\beta} (\phi, \chi, X_L)  + \rho_{\alpha\beta} (\phi)
+ \sigma_{\alpha\beta} (\chi)~,\\
\label{shiftambiguity2}   \bar{F}^+_{\alpha\beta} (\bar{\phi}, \bar{\chi}, 
\bar{X}_L) &\rightarrow &
\bar{F}^+_{\alpha\beta} (\bar{\phi}, \bar{\chi}, \bar{X}_L) + 
\bar{\rho}_{\alpha\beta}(\bar{\phi}) + \bar{\sigma}_{\alpha\beta} (\bar{\chi})~,\\
\label{shiftambiguity3}  F^-_{\alpha\beta}(\phi, \bar{\chi}, X_R) &\rightarrow &
F^-_{\alpha\beta}(\phi, \bar{\chi}, X_R) 
-\rho_{\alpha\beta}(\phi) -\bar{\sigma}_{\alpha\beta}(\bar{\chi})~,\\
\label{shiftambiguity4}  \bar{F}^-_{\alpha\beta} 
(\bar{\phi}, \chi, \bar{X}_R) &\rightarrow & \bar{F}^-_{\alpha\beta} 
(\bar{\phi}, \chi, \bar{X}_R) - \bar{\rho}_{\alpha\beta} 
(\bar{\phi}) - \sigma_{\alpha\beta}(\chi)~.
\eer{shiftambiguity} 
Here $\rho_{\alpha\beta}$ is holomorphic with respect to both $J_\pm$, and hence
is {\it biholomorphic}, and $\sigma_{\alpha\beta}$ is holomorphic with respect to
$J_+$ and $-J_-$ and hence is {\it twisted biholomorphic} (see the discussion below
in subsection \ref{biholcomment}).

From (\ref{doubleintersection}) we have the following relation
\beq
{\rm Re} (F^+_{\alpha\beta} + F^+_{\beta\alpha} + F^-_{\alpha\beta}
+ F^-_{\beta\alpha})=0~;
\eeq{symmetryoff}
using the ambiguities (\ref{shiftambiguity1})-(\ref{shiftambiguity4}) we can 
choose $F^+$'s and $F^-$'s to be separately antisymmetric under interchange of the open sets:
\beq
F^+_{\alpha\beta} = - F^+_{\beta\alpha}~,~~~~~~F^-_{\alpha\beta}=-F^-_{\beta\alpha}~.
\eeq{antisymmetry}

\subsection{Biholomorphic and twisted biholomorphic cocycles on triple overlaps}
Taking the coboundary of   (\ref{doubleintersection}) we get
on $U_\alpha \cap U_\beta \cap U_\gamma$
\beq
{\rm Re} (F^+_{\alpha\beta} + F^+_{\beta\gamma} + F^+_{\gamma\alpha} + 
 F^-_{\alpha\beta} + F^-_{\beta\gamma} + F^-_{\gamma\alpha})=0~,
\eeq{tripleintersection}
which implies that the coboundary of $F^\pm$ can be expressed in terms of biholomorphic
and twisted biholomorphic functions $c_{\alpha\beta\gamma}(\phi)$ and 
$b_{\alpha\beta\gamma}(\chi)$:
\ber
\label{bihol11}   F^+_{\alpha\beta} (\phi, \chi, X_L)+ 
F^+_{\beta\gamma} (\phi, \chi, X_L)+ F^+_{\gamma\alpha} (\phi, \chi, X_L)
 &=&  i\left(c_{\alpha\beta\gamma}(\phi) - b_{\alpha\beta\gamma}(\chi)\right)~,\\
\label{bihol22}   \bar{F}^+_{\alpha\beta} (\bar{\phi}, \bar{\chi}, \bar{X}_L)
+ \bar{F}^+_{\beta\gamma} (\bar{\phi}, \bar{\chi}, \bar{X}_L)+ \bar{F}^+_{\gamma\alpha} (\bar{\phi}, \bar{\chi}, \bar{X}_L) &=&  -i\left(\bar{c}_{\alpha\beta\gamma}(\bar{\phi}) 
- \bar{b}_{\alpha\beta\gamma}(\bar{\chi})\right)~,\\
\label{bihol33}   F^-_{\alpha\beta} (\phi, \bar{\chi}, X_R)
+ F^-_{\beta\gamma} (\phi, \bar{\chi}, X_R)+ F^-_{\gamma\alpha} (\phi, \bar{\chi}, X_R)
 &=&   - i\left(c_{\alpha\beta\gamma}(\phi) + \bar{b}_{\alpha\beta\gamma}(\bar{\chi})\right)~,\\
\label{bihol44}   \bar{F}^-_{\alpha\beta} (\bar{\phi}, \chi , \bar{X}_R)
+ \bar{F}^-_{\beta\gamma} (\bar{\phi}, \chi, \bar{X}_R)
+ \bar{F}^-_{\gamma\alpha} (\bar{\phi}, \chi, \bar{X}_R)
 &=&   i\left( \bar{c}_{\alpha\beta\gamma}(\bar{\phi}) 
+ b_{\alpha\beta\gamma}(\chi)\right)~,
\eer{resolutionoftriple}
where the $c$ and $b$ functions are defined up to constant shifts
\ber
\label{shiftincb1}  c_{\alpha\beta\gamma}(\phi) &\rightarrow& c_{\alpha\beta\gamma}(\phi) 
+ h_{\alpha\beta\gamma}~,\\
\label{shiftincb2} b_{\alpha\beta\gamma}(\chi) &\rightarrow& b_{\alpha\beta\gamma}(\chi) 
+ h_{\alpha\beta\gamma}~.
\eer{shiftincb}
At the same time, using (\ref{antisymmetry}) we can derive the relations
\ber
\label{symmetrycbgeneral1}   c_{\alpha\beta\gamma}(\phi) 
+ c_{\beta\alpha\gamma} (\phi) - b_{\alpha\beta\gamma}(\chi) -
b_{\beta\alpha\gamma} (\chi)&=& 0~,\\
\label{symmetrycbgeneral2}  c_{\alpha\beta\gamma}(\phi) 
+ c_{\beta\alpha\gamma} (\phi) + \bar{b}_{\alpha\beta\gamma}(\bar{\chi}) +
\bar{b}_{\beta\alpha\gamma} (\bar{\chi}) &= & 0~.
\eer{symmetrycbgeneral}
Using the ambiguities (\ref{shiftincb1})-(\ref{shiftincb2}) and the relations (\ref{symmetrycbgeneral1})-(\ref{symmetrycbgeneral2}) we can always choose such $c$ and $b$ which satisfy 
\beq
c_{\alpha\beta\gamma} = - c_{\beta\alpha\gamma}= - c_{\alpha\gamma\beta} 
= - c_{\gamma\beta\alpha}~,~~~~b_{\alpha\beta\gamma} = - b_{\beta\alpha\gamma}
= - b_{\alpha\gamma\beta} = - b_{\gamma\beta\alpha}~.
\eeq{antisymmetrycb}
\subsection{Integral cocycles on four-fold overlaps}
Now on $U_\alpha \cap U_\beta \cap U_\gamma \cap U_\delta$ we have
\ber
&& c_{\alpha\beta\gamma} + c_{\beta\alpha\delta}  + c_{\gamma\beta\delta} +
c_{\delta\alpha\gamma} - 
b_{\alpha\beta\gamma}  - b_{\beta\alpha\delta}  - b_{\gamma\beta\delta} -
b_{\delta\alpha\gamma} =0~,\\
&& c_{\alpha\beta\gamma} + c_{\beta\alpha\delta} + c_{\gamma\beta\delta} +
c_{\delta\alpha\gamma} + 
\bar{b}_{\alpha\beta\gamma} + \bar{b}_{\beta\alpha\delta}  
+ \bar{b}_{\gamma\beta\delta} + \bar{b}_{\delta\alpha\gamma}=0~.
\eer{cocyclecondition}
These conditions imply
\ber
\label{AB124}     && c_{\beta\gamma\delta} + c_{\delta\gamma\alpha} + c_{\alpha\beta\delta} +
c_{\beta\alpha\gamma} = \frac {i}{4} d_{\alpha\beta\gamma\delta}~,\\
\label{AB1245}   &&   b_{\beta\gamma\delta} + b_{\delta\gamma\alpha} + b_{\alpha\beta\delta} +
b_{\beta\alpha\gamma} =  \frac {i}{4} d_{\alpha\beta\gamma\delta}~.
\eer{resolutiongerbemmms}
In particular using the formulas in  \cite{Lindstrom:2005zr}  for $H$ in terms of $K$ we see that if $\frac{H}{2\pi}
\in H^3(M, \mathbb{Z})$ then $ d_{\alpha\beta\gamma  \delta} \in 2\pi \mathbb{Z}$. 

Let us elaborate on the relation between the generalized \kah potential and 
the description of a gerbe with   connection from Section \ref{holomorphicgerbe}. 
The key observation is that  using   formulas from \cite{Lindstrom:2005zr} we can find 
a {\em linear} expression for  the one-form potentials $\lambda_\pm$  (see (\ref{omegalambda})) in terms of the
generalized K\"ahler potential as
\ber
\label{AAABBB1}
\lambda_+^{(1,0)}+\bar\lambda_+^{(0,1)} &=& 
-i\left(\frac{\partial K}{\partial X_R^{\alpha'}} 
dX_R^{\alpha'}+
\frac{\partial K}{\partial \phi^a}d\phi^a -
\frac{\partial K}{\partial \chi^{a'}}d\chi^{a'}
\right) - c.c.
\\
\nonumber&=& -\left(\frac{\partial K}{\partial X_R^{\cal A'}} 
(J_-)^{\cal A'}_{\;\;\; \cal B'}dX_R^{\cal B'}+
\frac{\partial K}{\partial \phi^A}(J_-)^A_{\;\;\; B}d\phi^B+
\frac{\partial K}{\partial \chi^{A'}}(J_-)^{A'}_{\;\;\; B'}d\chi^{B'}
\right)~,
\\
\nonumber\lambda_-^{(1,0)} +\bar\lambda_-^{(0,1)}&=& 
i\left(\frac{\partial K}{\partial X_L^{\alpha}} 
dX_L^{\alpha}+
\frac{\partial K}{\partial \phi^a}d\phi^a+
\frac{\partial K}{\partial \chi^{a'}}d\chi^{a'}
\right) - c.c.
\\
\label{AAABBB2}&=& \left(\frac{\partial K}{\partial X_L^{\cal A}} 
(J_+)^{\cal A}_{\;\;\; \cal B}dX_L^{\cal B}+
\frac{\partial K}{\partial \phi^A}(J_+)^A_{\;\;\; B}d\phi^B+
\frac{\partial K}{\partial \chi^{A'}}(J_+)^{A'}_{\;\;\; B'}d\chi^{B'}
\right)~,
\eer{aaabbbtotal}
(see  \cite{Lindstrom:2005zr}  for our index conventions.)
Then comparing with (\ref{doubleholgerbe}), we find that in 
the double overlap 
$ {  U}_\alpha \cap {  U}_\beta$ 
\ber
\xi_+ &=& i(\bar F^- - F^-)
\\
\phi_+^{(1,0)} &=& i(\frac{\partial F^+}{\partial\chi^{a'}} d\chi^{a'} 
- \frac{\partial F^+}{\partial \phi^a} d\phi^a)
\\
\xi_- &=& i(F^+ - \bar F^+)
\\
\phi_-^{(1,0)} &=& i(\frac{\partial F^-}{\partial\phi^a} d\phi^a 
- \frac{\partial F^-}{\partial\bar\chi^{\bar a'}} d\bar\chi^{\bar a'})
\eer{coolformulaswhichshouldbenamed}
from which follows (using (\ref{definitionLambdaholgerbe}) and (\ref{bihol11})--(\ref{bihol44}))
\beq
\Lambda_{\alpha\beta\gamma} = i\left(\bar c_{\alpha\beta\gamma}(\bar{\phi}) 
- {c}_{\alpha\beta\gamma}(\phi)
+ \bar b_{\alpha\beta\gamma} (\bar{\chi}) - {b}_{\alpha\beta\gamma}(\chi)\right).
\eeq{solutionforlambdafromK}
Assuming the integrality of $H$ and the relations (\ref{AB124}, \ref{AB1245}) we find 
that (\ref{chainofrelations4}) is satisfied with $ d_{\alpha\beta\gamma  \delta} \in 2\pi \mathbb{Z}$. 
This allows  us to introduce the functions
\beq
G_{\alpha\beta\gamma} (\phi) = e^{4c_{\alpha\beta\gamma}(\phi)}~,~~~~~~
F_{\alpha\beta\gamma} (\chi)= e^{4 b_{\alpha\beta\gamma}(\chi)}~,
\eeq{gerbdefiniion}
defined over triple intersections 
\beq
G_{\alpha\beta\gamma}~, F_{\alpha\beta\gamma}~:~U_\alpha\cap 
U_\beta \cap U_\gamma~,\rightarrow~\mathbb{C}^*~,
\eeq{definitionbiholgerbe}
which are antisymmetric under permutations of the open sets
and satisfy the cocycle condition on the four-fold intersection.

\subsection{Comments on biholomorphic functions}
\label{biholcomment}
Note that $G$ depends only on $\phi$ and $F$ depends only on $\chi$. 
On any manifold $M$ with two complex structures $J_+$ and $J_-$, 
a complex function $f$ is biholomorphic if it is 
holomorphic with respect to both complex structures, \ie
\beq
(d- i d^c_\pm) f = 0~.
\eeq{definitionofbihol}
As a result of this $f$ also satisfies
\beq
(d^c_+ - d^c_-) f =0~.
\eeq{relaywi299912}
Thus on a generalized \kah manifold a biholomorphic 
function $f$ is a Casimir function of 
the Poisson structure $\pi_-$, \ie $df \in \ker \pi_-$.  
In the coordinate system we use, this means that 
a biholomorphic function is a function of only $\phi$. 
Analogously one can consider functions which are 
holomorphic with respect to $J_+$ and antiholomorphic 
with respect to $J_-$, which we call twisted 
biholomorphic. Such functions are the Casimir 
functions for $\pi_+$ and in our coordinates 
depend only on $\chi$. 

\subsection{Biholomorphic gerbes}
Combining the above, we arrive at the definition of 
a biholomorphic gerbe on a bicomplex 
manifold $(M, J_+, J_-)$ as a collection of 
biholomorphic functions $G_{\alpha\beta\gamma}$ 
that are antisymmetric under permutations of the open sets and
satisfy the cocycle condition on four-fold intersections. 
Analogously we define a twisted biholomorphic gerbe as 
a collection of twisted biholomorphic functions $F_{\alpha\beta\gamma}$.
In the analysis above we have shown 
that a  bihomolomorphic gerbe and a   twisted bihomolomorphic gerbe  naturally 
appear in the context of generalized \kah geometry through the discussion 
of gluing of the generalized \kah potential.  Moreover, by exponentiating the equations
(\ref{bihol11}-\ref{bihol44}) we arrive at the following relation between 
biholomorphic and twisted  biholomorphic transition  functions 
\beq
G_{\alpha\beta\gamma}  F^{-1}_{\alpha\beta\gamma}
= h^+_{\alpha\beta} h^+_{\beta\gamma} h^+_{\gamma\alpha}~,~~~~~~~~~~~~~~
G_{\alpha\beta\gamma} \bar{F}_{\alpha\beta\gamma}
= h^-_{\alpha\beta} h^-_{\beta\gamma} h^-_{\gamma\alpha} ~,
\eeq{relationbetweengege}
where $h^\pm_{\alpha\beta} = exp (\mp 4i F^\pm_{\alpha\beta})$ 
are $J_\pm$-holomorphic functions of   special form. 

Alternatively, one can define this structure using transition line bundles; essentially,
one defines a holomorphic transition line bundle for each complex structure on the 
generalized K\"ahler manifold $M$ and imposes a compatibility condition which is
equivalent to (\ref{relationbetweengege}). This description is sometimes more compact,
and is used in the example of a biholomorphic gerbe on $S^1\times S^3$ given in the
Appendix.

In the special case in which $(J_+ - J_-)$ is invertible everywhere on $M$, 
then the only biholomorphic functions are constants.  Indeed in this 
case $H$ is cohomologically trivial\footnote{This is easy to see in 
the formulation of generalized \kah geometry in terms of generalized 
complex structures. In this case one of the generalized complex structures 
is of symplectic type and therefore $H$ is exact.}. 
Analogously one can show that if $(J_+ + J_-)$ is invertible everywhere on $M$, then
the only twisted biholomorphic functions are constants and again $H$ is exact.

The present analysis is not complete. We have focused on the linear 
transformations (\ref{doubleintersection}) of $K$; however,
as shown in \cite{Lindstrom:2005zr}, $K$ also encodes a 
choice of polarization for symplectic leaves on the manifold. 
A change of this polarization is realized by a Legendre transformation, and
we have not explored how this nonlinear transformation intertwines with the linear
transitions discussed above.
We leave the
problem of finding a full geometrical interpretation of $K$ for future research.

 \Section{Bihermitian Local Product spaces}
\label{BILP}

In this Section we consider the special case of regular generalized \kah manifolds
for which $\sigma = [J_+, J_-]g^{-1}=0$. This means that we can   simply exclude $X_{L,R}$ and $\bar{X}_{L,R}$ from all formulas 
of the previous Section.   
This case is considerably simpler than the general case since all 
geometrical objects depend linearly on the generalized \kah potential
and it is guaranteed that every point is regular. Such a space carries a local product
structure $\Pi = J_+ J_-$ and is called a Bihermitian Local Product space (or BiLP for short). 
Locally, a BiLP looks like a product of two \kah manifolds and in a way $H$ is responsible 
for making this product nontrivial. The simplest compact example of such geometry is 
$S^3 \times S^1$, and this example is analyzed in the Appendix.

\subsection{The generalized \kah potential on a BiLP}
On a BiLP, the complex structures $J_\pm$ commute, and hence
the differentials $d^c_\pm$ obey
\beq
d_+^c d_-^c = - d_-^c d_+^c
\eeq{dcpdcm}
as well as $d d_\pm^c = - d_\pm^c d$. Hence, on a BiLP we can write
the closed three-form $H$ as
\beq
H = \frac12 d d_+^c d_-^c K_\alpha~,
\eeq{localformHforBILP}
where $K_\alpha$ is a real function on a patch $U_\alpha$ 
and $H$ is $(2,1)+(1,2)$ form with respect 
to both complex structures $J_\pm$.  Comparing (\ref{localformHforBILP}) 
and (\ref{definitionofH}), we see that the two-forms $\omega_\pm$ can also be simply 
expressed in terms of $K_\alpha$. 

There are a number of compatible distributions on $TM$ given by $J_+$, $J_-$ and $\Pi$.
This allows us to have quadruple-grading on the differential forms   
and split the de Rham differential as follows
\beq
d = \partial_\phi + \partial_\chi + \partial_{\bar\phi} + \partial_{\bar\chi}~. 
\eeq{blablabla}
Further details on the geometry and notation can be found in \cite{Gates:1984nk},\cite{Lindstrom:2007qf}. 

\subsection{Gerbes on a BiLP}
It is a straightforward exercise  to work out the whole chain of the 
relations (\ref{coB11form}-\ref{doubleholgerbe})
and (\ref{chainofrelationshol1}-\ref{chainofrelationshol3}) in
terms of the data coming from the gluing of the generalized \kah
potential $K_\alpha$. As particular examples, let us present the 
following expressions (also see the equations
(\ref{AAABBB1}-\ref{coolformulaswhichshouldbenamed}) 
from the previous Section). Choose, \eg $J_+$ as
the complex structure  with respect to which the differential forms are graded;
the relations (\ref{chainofrelationshol1}-\ref{chainofrelationshol3}) are satisfied by 
\ber
B^{(2,0)}_{\alpha} &=& 2\left(\partial_{\bar\phi} \partial_{\bar\chi}
+ \partial_\phi \partial_\chi\right)K_{\alpha}~,\\
A^{(1,0)}_{\alpha\beta} &=& -2\partial_\phi F^+_{\alpha\beta}~,\\    
G_{\alpha\beta\gamma} &=& e^{4c_{\alpha\beta\gamma}}~,
\eer{ddkk3030222}
where the transition functions are chosen to be biholomorphic. 
Indeed, one can write many more formulas 
along these lines which would correspond to different but equivalent ways
of describing the gerbe with  connection. There {\em always} 
exists a choice that makes the transition functions (twisted) biholomorphic. 

In the case of a holomorphic line bundle the existence of a Hermitian structure
(\ref{hermitianlinebundle}) is equivalent to 
the existence of a \kah potential.  However, in the context of a holomorphic gerbe, 
a Hermitian structure only implies the existence of real function $h_{\alpha\beta}$
on double intersections, see (\ref{definitionhermitiangerbe}).
These functions $h_{\alpha\beta}$ are interpreted as exponents of the \kah potential 
for the transition line bundle (see the discussion at the end of Section \ref{realgerbe}).
The holomorphic gerbe with Hermitian structure does not naturally 
produce a real function defined over a patch $U_\alpha$.
However  in the present context we can introduce
the notion of a bihermitian structure on a (twisted) biholomorphic gerbe that is 
equivalent to the existence of the generalized \kah potential. 

Suppose that on a bicomplex manifold $(M, J_+, J_-)$ 
we have transition functions for a biholomorphic gerbe 
$G_{\alpha\beta\gamma}$ and transition
functions for a twisted biholomorphic gerbe $F_{\alpha\beta\gamma}$.  
The biholomorphic and twisted biholomorphic gerbes 
are are both hermitian if the following condition is satisfied:
\beq
G_{\alpha\beta\gamma}  F^{-1}_{\alpha\beta\gamma}= h^+_{\alpha\beta} 
h^+_{\beta\gamma} h^+_{\gamma\alpha}~,~~~~~~~~~~~~~~
G_{\alpha\beta\gamma} \bar{F}_{\alpha\beta\gamma}= 
h^-_{\alpha\beta} h^-_{\beta\gamma} h^-_{\gamma\alpha} ~,
\eeq{bliadi-ia-ustal}
where $h^\pm_{\alpha\beta}$ are $J_\pm$-holomorphic functions 
on double intersections.  The products $G_{\alpha\beta\gamma} 
F_{\alpha\beta\gamma}$ are $J_+$-holomorphic functions on the triple 
intersections which satisfy the cocycle condition on the four-fold intersections, 
since both $G_{\alpha\beta\gamma}$ and $F_{\alpha\beta\gamma}$ 
satisfy the cocycle conditions independently.   
Moreover, from (\ref{bliadi-ia-ustal}) the product 
$(GF\bar{G}\bar{F})_{\alpha\beta\gamma}$ 
is a real trivial cocycle. Therefore  $G_{\alpha\beta\gamma} F_{\alpha\beta\gamma}$ 
can be interpreted as the transition functions for a $J_+$-holomorphic gerbe 
with Hermitian structure. Analogously,  $G_{\alpha\beta\gamma} 
\bar{F}^{-1}_{\alpha\beta\gamma}$ can be interpreted as the 
transition functions for a $J_-$-holomorphic gerbe
with Hermitian structure. 

Furthermore,  (\ref{bliadi-ia-ustal}) implies the condition 
\beq
h^+_{\alpha\beta} h^+_{\beta\gamma} h^+_{\gamma\alpha} \bar{h}^-_{\alpha\beta} 
\bar{h}^-_{\beta\gamma} \bar{h}^-_{\gamma\alpha} 
= h^-_{\alpha\beta} 
h^-_{\beta\gamma} h^-_{\gamma\alpha} 
\bar{h}^+_{\alpha\beta} \bar{h}^+_{\beta\gamma} 
\bar{h}^+_{\gamma\alpha} 
~,
\eeq{condisusuwiioqoq}
which implies $h^+(\bar h^+ )^{-1}(h^-)^{-1}\bar h^-$ is a trivial  real cocycle:
\beq
h^+_{\alpha\beta} \bar{h}^-_{\alpha\beta} 
( h^-_{\alpha\beta})^{-1} ( \bar{h}^+_{\alpha\beta})^{-1} =
e^{K_\alpha} e^{-K_\beta}~,
\eeq{finallypotential}
where the real function $K_\alpha$ is defined on $U_\alpha$ and can be interpreted as 
a generalized \kah potential. Thus, on a bihermitian local product space, given a 
biholomorphic and twisted biholomorphic gerbe satisfying the bihermitian 
compatibility condition (\ref{bliadi-ia-ustal}), one can always construct a 
generalized K\"ahler potential.  

 \Section{Summary and Conclusions}
\label{summary}

We have discussed aspects of gerbes that arise naturally on generalized \kah geometries.
These geometries are bihermitian, with two complex structures and a related single
closed three-form $H$. It is natural to construct two holomorphic gerbes with the same
curvature $H$. The additional structure of the generalized \kah geometries allows
one to describe them in terms of a generalized \kah potential (away from irregular points).
Using this potential, we showed that the two gerbes fit together into a structure we
called the biholomorphic gerbe, whose transition functions can be chosen to be holomorphic
with respect to both complex structure (biholomorphic). When the complex structures
commute, we were able to explicitly reverse the construction and use the gerbe to
construct the generalized \kah potential. We believe that this should be possible in general,
and that the biholomorphic gerbe plays the same role for generalized \kah geometry as the
holomorphic line bundle plays for \kah geometry.

The generalized  (K\"ahler) potential arises naturally as the superspace 
Lagrange density in the sigma model approach \cite{Lindstrom:2005zr}; 
a geometric interpretation as a generator of symplectomorphisms 
related to a choice of polarization was given for the case in which 
the Poisson structures of the manifold are nondegenerate. 
Here we see that in the complementary (BiLP) case, when the
Poisson structures are maximally degenerate, it has an 
interpretation as a new kind of bihermitian structure on a 
biholomorphic gerbe. It would be very interesting to understand 
the general case, which should combine both perspectives. In particular, the global 
characterization of the generalized K\"ahler potential should allow for changes of 
polarization generated by Legendre transforms, as well as generalized K\"ahler 
gauge transformations in the transitions between coordinate patches.

Other extensions of this work that are of immediate interest concern irregular
points. Generically, these form (real) co-dimension two loci within the manifold,
and can carry nontrivial topology -- this corresponds to charge for the gerbe
connection. 

In the Appendix, we give a careful discussion of the generalized \kah structure
on $S^3\times S^1$ viewed as a BiLP.  There are other 
generalized \kah structures on $S^3\times S^1$ that exemplify type change; these
will be presented in a separate publication.

\bigskip\bigskip
\noindent{\bf\Large Acknowledgement}:
\bigskip\bigskip

\noindent
We thank Gil Cavalcanti, Ezra Getzler, Marco Gualtieri, 
Nigel Hitchin, Blaine Lawson, and John Morgan
for discussions. We are grateful to the 2007 and 2008 
Simons Workshop for providing the stimulating atmosphere
where part of this work was carried out. We thank 
the program ``Poisson sigma models, Lie algebroids, 
deformations and higher analogues"
at the Erwin Schr\"odinger International Institute for Mathematical Physics 
where part of this work was carried out. We are particularly happy to
thank the program ``Geometrical Aspects of String Theory'' at Nordita, 
where this work was finished.
The research of UL was supported by EU grant (Superstring theory)
MRTN-2004-512194 and VR grant 621-2006-3365.
The research of MR  was supported in part by NSF grant no.~PHY-06-53342.
The research of R.v.U. was supported by
the Simons Center for Geometry and Physics as well as the
Czech ministry of education under contract No.~MSM0021622409.
The research of M.Z. was supported by VR-grant 621-2004-3177.

\appendix
 \Section{Appendix: $S^3\times S^1$}
\setcounter{equation}{0}
The manifold $S^3\times S^1$ is a well known example that illustrates many aspects
of the previous discussion. It has a bihermitian structure \cite{rss,ikr}
with a nontrivial biholomorphic gerbe. The curvature of the gerbe is proportional
to the volume form of the $S^3$ factor; classically, we can normalize it as we choose,
but in the  quantum theory, the normalization determines the level of the corresponding
WZW-model. 

The manifold has a Bihermitian Local Product (BiLP) structure, and in adapted
complex coordinates $\phi,\chi$, the metric can be written as:
\be\label{sxsds}
ds^2=\frac{1}{8\pi}
\left[\frac{d\phi d\bar\phi+ d\chi d\bar\chi}{\phi\bar\phi+\chi\bar\chi}\right]~.
\ee
Here the direction corresponding to the homothety (uniform rescaling of $\phi,\chi$)
lies along the $S^1$ factor, and the $S^3$ is found by fixing $\phi\bar\phi+\chi\bar\chi$
to a constant. We compactify the homothetic coordinate to $S^1$ by 
the restriction 
\be\label{slrange}
1\le\phi\bar\phi+\chi\bar\chi<e^{4\pi}~.
\ee

The metric can be given a  more familiar form by introducing real coordinates
\be\label{realcoor}
\phi  &=& {\rm e}^{r+i\varphi} \sin\theta~,\\
\chi &=&   {\rm e}^{r+i\psi}\cos\theta~,
\ee
in which the metric becomes
\be
ds^2 = \frac{1}{8\pi}
\left(dr^2 + d\theta^2 + \sin^2\theta d\varphi^2 + \cos^2\theta d\psi^2\right)~.
\ee
The $r$ direction decouples and corresponds to the $S^1$. The $S^3$ is described as
an interval ($0\leq \theta\leq \frac{\pi}{2}$) with a torus fibration over it (the torus being
coordinatized by $0\leq \varphi,\psi \leq 2\pi$). The torus degenerates at the endpoints
of the interval.

The metric (\ref{sxsds}) can integrated to give a generalized K\"ahler potential \cite{rss}
\be
K_\alpha = \frac{1}{8\pi}\left[\frac12 \left(\ln\phi\bar\phi\right)^2
-\int\limits_1^\frac{\chi\bar\chi}{\phi\bar\phi} dx \;\frac{\ln(1+x)}{x}\right]~.
\ee
This is well defined in a patch ${\cal U}_\alpha$ where $\phi \neq 0$. When $\phi$ goes
to zero we can make a K\"ahler gauge transformation and go to a generalized K\"ahler
potential
\be
K_\beta = \frac{1}{8\pi}\left[ -\frac12 \left(\ln\chi\bar\chi\right)^2
+\int\limits_1^\frac{\phi\bar\phi}{\chi\bar\chi} dx \;\frac{\ln(1+x)}{x}\right]~,
\ee
which is well defined in the open set ${\cal U}_\beta$ where $\chi\neq 0$. The K\"ahler
gauge transformation taking us between ${\cal U}_\alpha$ and ${\cal U}_\beta$ is
\be
K_\alpha - K_\beta = \frac{1}{8\pi}\left[\ln\!\left(\phi\bar\phi\right)
\ln\!\left(\chi\bar\chi\right)\right]~,
\ee
which also immediately tells us that, {\it cf.} (\ref{doubleintersection})
\be
F^+_{\alpha\beta} &=& \frac{1}{8\pi}\ln\phi\ln\chi~,\\
F^-_{\alpha\beta} &=& \frac{1}{8\pi}\ln\phi\ln\bar\chi~.
\ee

This is enough information to define the gerbe for this 
generalized K\"ahler manifold. The simplest definition in this 
case is in terms of the ``transition line bundle" on double overlaps
as summarized at the end of section 6. The double overlap 
${\cal U}_{\alpha\beta} = {\cal U}_\alpha \cap {\cal U}_\beta$ 
has the topology of a cylinder (an interval times the
$S^1$ factor) times a torus (constructed from the 
phases of $\phi$ and $\chi$). The gerbe is specified by 
giving a holomorphic line bundle on this set. The first Chern class of 
this bundle is specified by an integer, which determines the relative factor 
between the volume form on $S^3$ and the Chern class of the gerbe 
(and gives the level of the corresponding WZW model).

We can check that this interpretation makes sense since 
the information that we have is enough to compute the K\"ahler 
form of this line bundle as well as the connection one-form
and the transition functions. Using the data we have
\be
\omega = 2\partial\bar\partial\left(\bar F^- - F^-\right) &=& 
\frac{1}{4\pi}\left(\frac{d\chi}{\chi}\wedge\frac{d\bar\phi}{\bar\phi}
-\frac{d\phi}{\phi}\wedge\frac{d\bar\chi}{\bar\chi}\right)~,\\
A^{(1,0)} &=& \frac{1}{8\pi}\left[
\ln\bar\chi\; \frac{d\phi}{\phi}-\ln\bar\phi\;\frac{d\chi}{\chi}\right]~,
\ee
which, if we use the real coordinates (\ref{realcoor})
becomes
\be
\omega = \frac{1}{2\pi}\left( \frac{dr\wedge d\theta}{\sin\theta\cos\theta}
+d\psi\wedge d\varphi\right)~.
\ee
We can also compute $H$:
\be
H =\frac{1}{\pi}\sin\theta\cos\theta d\theta\wedge d\varphi \wedge d\psi~,
\ee
which is the volume form of $S^3$ with the normalization $2\pi$ so that $\frac{H}{2\pi}$
is integral.

We could now go on to find the transition functions of the 
gerbe on triple overlaps. To do this we would first
need to subdivide ${\cal U}_\alpha$ and ${\cal U}_\beta$ 
to define a good cover. The nontrivial transition functions 
are then associated with the phases one picks up
going around the torus defined from the phases of $\phi$ and $\chi$.


\begin{thebibliography}{6666}

\newcommand{\np}{{\em Nucl.\ Phys.\ }}
\newcommand{\pr}{{\em Phys.\ Rev.\ }}
\newcommand{\cmp}{{\em Commun.\ Math.\ Phys.\ }}
\newcommand{\pl}{{\em Phys.\ Lett.\ }}
%
\bibitem{gualtieriPhD}
M.~Gualtieri, {\it Generalized complex geometry,} Oxford University  
DPhil thesis, [arXiv:math.DG/0401221];\\
{\it Generalized complex geometry,}  [arXiv:math.DG/0703298].
%
\bibitem{Gates:1984nk}
S.~J.~Gates, C.~M.~Hull and M.~Ro\v{c}ek,
{\it Twisted Multiplets And New Supersymmetric Nonlinear Sigma Models,}
Nucl.\ Phys.\  B {\bf 248} (1984) 157.
%
\bibitem{Howe:1985pm}
P.~S.~Howe and G.~Sierra,
{\it Two-Dimensional Supersymmetric Nonlinear Sigma Models With Torsion},
Phys.\ Lett.\  B {\bf 148}, 451 (1984).
%
\bibitem{Buscher:1987uw}
T.~Buscher, U.~Lindstr\"om and M.~Ro\v cek,
{\it New Supersymmetric Sigma Models with Wess-Zumino Terms},
Phys.\ Lett.\  B {\bf 202}, 94 (1988).
%
\bibitem{Sevrin:1996jr}
A.~Sevrin and J.~Troost,
{\it Off-shell formulation of N = 2 nonlinear sigma models},
Nucl.\ Phys.\  B {\bf 492}, 623 (1997)
[arXiv:hep-th/9610102].
%
\bibitem{Lindstrom:2005zr}
U.~Lindstr\"om, M.~Ro\v{c}ek, R.~von Unge and M.~Zabzine,
{\it Generalized Kaehler manifolds and off-shell supersymmetry,}
Commun.\ Math.\ Phys.\  {\bf 269} (2007) 833
[arXiv:hep-th/0512164].
%
\bibitem{Lindstrom:2007xv}
U.~Lindstr\"om, M.~Ro\v{c}ek, R.~von Unge and M.~Zabzine,
{\it A potential for generalized Kaehler geometry,}
arXiv:hep-th/0703111.
%

%
\bibitem{Gawedzki:2002se}
  K.~Gawedzki and N.~Reis,
  Rev.\ Math.\ Phys.\  {\bf 14}, 1281 (2002)
  [arXiv:hep-th/0205233].
  
\bibitem{Gawedzki:2004tu}
  K.~Gawedzki,
  {\it Abelian and non-Abelian branes in WZW models and gerbes,}
  Commun.\ Math.\ Phys.\  {\bf 258}, 23 (2005)
  [arXiv:hep-th/0406072].
 
 \bibitem{Carey:2004xt}
  A.~L.~Carey, S.~Johnson, M.~K.~Murray, D.~Stevenson and B.~L.~Wang,
  {\it Bundle gerbes for Chern-Simons and Wess-Zumino-Witten theories,}
  Commun.\ Math.\ Phys.\  {\bf 259}, 577 (2005)
  [arXiv:math/0410013].
 
\bibitem{Aschieri:2004yz}
  P.~Aschieri and B.~Jurco,
  {\it Gerbes, M5-brane anomalies and E(8) gauge theory,}
  JHEP {\bf 0410}, 068 (2004)
  [arXiv:hep-th/0409200].

\bibitem{Schreiber:2005mi}
  U.~Schreiber, C.~Schweigert and K.~Waldorf,
  {\it Unoriented WZW models and holonomy of bundle gerbes,}
  Commun.\ Math.\ Phys.\  {\bf 274}, 31 (2007)
  [arXiv:hep-th/0512283].

\bibitem{Belov:2007qj}
  D.~M.~Belov, C.~M.~Hull and R.~Minasian,
  {\it T-duality, Gerbes and Loop Spaces,}
  arXiv:0710.5151 [hep-th].
%
  
\bibitem{hitchinwhat}
N.~Hitchin, {\it What is a Gerbe?}, Notices Amer.\ Math.\ Soc.\ , {\bf 50} (2), 218-219.
%
\bibitem{hitchingerbe}
N.~Hitchin, {\it Lectures on special Lagrangian submanifolds,} 
in ``Winter School on Mirror symmetry, Vector Bundles and 
Lagrangian Submanifolds,'' C.~Vafa and S.-T.~Yau (eds.), Studies in Advanced 
Mathematics 23, AMS/International Press, Providence (2001), 151-182,
arXiv:math/9907034.
%
\bibitem{woodhouse}
N.~M.~J.~Woodhouse,  {\it Geometric quantization},
Second edition. Oxford Mathematical Monographs. Oxford Science Publications. 
The Clarendon Press, Oxford University Press, New York, {\bf 1992}. 
%
\bibitem{giraud}
J.~Giraud, {\it Cohomologie nonab\'elienne,}
Grundl. {\bf 179}, Springer Verlag, Berlin (1971).
%
\bibitem{brylinski}
J.-L.~Brylinski, {\it Loop spaces, Characteristic Classes and Geometric Quantization,}
Progr. Math. {\bf 107}, Birkh\"auser, Boston-Basel, 1993.
%
\bibitem{chatterjee}
D.~S.~Chatterjee, {\it On gerbs}, PhD thesis, Cambridge, 1998.
%
\bibitem{grundle}
B.~Lawson and F.~Reese~Harvey,
{\it From Sparks to Grundles - Differential Characters,}
Comm. in Anal. and Geom. vol. 14 (2006), 1-34
%
\bibitem{murray}
M.~K.~Murray, 
{\it An Introduction to Bundle Gerbes,}
arXiv:0712.1651 [math.DG].

\bibitem{Hitchin:1986ea}
N.~J.~Hitchin, A.~Karlhede, U.~Lindstr\"om and M.~Ro\v{c}ek,
{\it Hyperkahler Metrics and Supersymmetry,}
Commun.\ Math.\ Phys.\  {\bf 108} (1987) 535.
%
\bibitem{Hull:1985jv}
C.~M.~Hull and E.~Witten,
{\it Supersymmetric Sigma Models And The Heterotic String},
Phys.\ Lett.\  B {\bf 160}, 398 (1985).
%
\bibitem{Hull:2008vs}
C.~M.~Hull, U.~Lindstr\"om, L.~Melo dos Santos, R.~von Unge and M.~Zabzine,
{\it Topological Sigma Models with H-Flux,}
JHEP {\bf 0809}, 057 (2008)
[arXiv:0803.1995 [hep-th]].
%
\bibitem{poisson-gualtieri}
M.~Gualtieri, {\it Branes on Poisson varieties,}
[arXiv:0710.2719]. 
%
\bibitem{Lyakhovich:2002kc}
S.~Lyakhovich and M.~Zabzine,
{\it Poisson geometry of sigma models with extended supersymmetry,}
Phys.\ Lett.\  B {\bf 548} (2002) 243
[arXiv:hep-th/0210043].
%
\bibitem{Hitchin:2005cv}
N.~Hitchin,
{\it Instantons, Poisson structures and generalized Kaehler geometry},
Commun.\ Math.\ Phys.\  {\bf 265} (2006) 131
[arXiv:math/0503432].
%
\bibitem{Lindstrom:2007qf}
U.~Lindstr\"om, M.~Ro\v{c}ek, R.~von Unge and M.~Zabzine,
{\it Linearizing Generalized Kahler Geometry,}
JHEP {\bf 0704} (2007) 061
[arXiv:hep-th/0702126].
%
\bibitem{rss}
M.~Ro\v{c}ek, K.~Schoutens and A.~Sevrin,
{\it Off-Shell WZW Models In Extended Superspace},
Phys.\ Lett.\  B {\bf 265}, 303 (1991).
%
\bibitem{ikr}
I.~T.~Ivanov, B.~b.~Kim and M.~Ro\v{c}ek,
{\it Complex structures, duality and WZW models in extended superspace},
Phys.\ Lett.\  B {\bf 343}, 133 (1995)
[arXiv:hep-th/9406063].



\end{thebibliography}
\end{document}